\documentstyle[aps,prl,epsf]{revtex}
\begin{document}
\draft

\title{General transport properties of superconducting quantum point
contacts: a Green functions approach}

\author{A. Martin-Rodero, A. Levy Yeyati, J.C. Cuevas}
\address{Departamento de F\'\i sica Te\'orica de la Materia Condensada C-V.\\
Facultad de Ciencias. Universidad Aut\'onoma de Madrid.\\
E-28049 Madrid. Spain.}

\date{\today}
\maketitle

\begin{abstract}
We discuss the general transport properties of
superconducting quantum point contacts. We show how these properties 
can be obtained from a microscopic model using nonequilibrium Green 
function techniques. 
For the case of a one-channel contact we analyze the response under
different biasing conditions: constant applied voltage, current
bias and microwave-induced transport. Current fluctuations are
also analyzed with particular emphasis on thermal and shot-noise.
Finally, the case of superconducting transport through a resonant 
level is discussed. The calculated properties show a remarkable
agreement with the available experimental data from atomic-size
contacts measurements. We suggest the possibility of extending this 
comparison to several other predictions of the theory. 
\end{abstract}

\section{Introduction}
Since the discovery of the Josephson effect \cite{Joseph} the
electronic transport between weakly coupled superconducting electrodes
(weak superconductivity) has been a subject of growing interest
\cite{Tinkham}.
Typically, weak superconductivity has been studied in
SIS, SNS and S-c-S junctions, where S, N, I and c denote superconductor,
normal metal, insulator and constriction respectively.
Recent technological advances have made possible the fabrication of
mesoscopic S-c-S junctions in which the electrodes are connected by 
a small number of conduction channels. These systems are usually referred to as
superconducting quantum point contacts (SQPC), examples of which
are the S-2DEG-S junctions \cite{Takayanagi} and
atomic contacts produced by break junctions \cite{Post,Scheer1} and
scanning tunneling microscope (STM) \cite{Scheer2} techniques.

On the theoretical side there has also been a parallel advance with the
development of fully quantum mechanical theories for the transport
properties of superconducting one-channel contacts 
\cite{Averin1,tocho,Hurd,Bratus}.
There has been a remarkable agreement between theoretical 
predictions and experimental results 
for the quantities that have so far been measured.
These quantities include the phase-dependent supercurrent in a
high transmissive contact \cite{Koops} and the dc current at constant
bias voltage \cite{Scheer1,Scheer2}. As we discuss in this paper, there remain
many exciting predictions of the microscopic theories to be
explored experimentally.

The aim of this paper is to present an overview of the main theoretical
results that have been obtained for different microscopic models of an
SQPC. An interesting aspect of superconducting transport is that
qualitatively different behaviors are exhibited depending on how the
system is biased. This will be analyzed in this work by
discussing the cases of phase, voltage and current bias together with
the case of transport under microwave radiation.
The models are introduced in section II together with the
nonequilibrium Green functions formalism used to calculate their
transport properties. Section III is
devoted to the voltage biased case for which we discuss the comparison
of the fully quantum mechanical calculation with semiclassical standard
theories and the available experimental results. We also discuss the
limit of very small voltage. In section IV the
current biased case is briefly analyzed while the response under
microwave radiation and its possible relevance for directly detecting
Andreev states is discussed in section V. Thermal and shot-noise are
the subject of section VI where we discuss the conditions
for observing coherent transport of multiple charge quanta from the
noise-current ratio. Finally in section VII the
superconducting transport through a resonant level is analyzed
both in the limits of very
large and very small charging energy.
The general conclusions are summarized in section VIII.

\section{Microscopic model and Green function formalism}

A schematical representation of a quantum point contact is depicted in
Fig. 1. For a typical point contact the length of the constriction
between the electrodes, $L_C$, is much smaller than the superconducting
coherence length $\xi_0$ and its width $W_C$ is $\sim \lambda_F$, the
electron Fermi wavelength. The first condition ensures that the detailed
superconducting phase and electrochemical potential profiles in the
constriction region are unimportant and can be safely approximated by
step functions. On the other hand, the condition $W \sim \lambda_F$
implies that there are only a few conduction channels between the
electrodes.

\begin{figure}[!htb]
\begin{center}
$\,$
\vspace{1cm}
\epsfxsize=5cm
\epsfbox{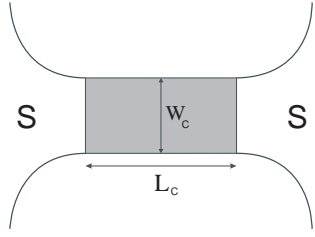}
\caption{Schematic representation of a superconducting quantum point
contact.}
\end{center}
\end{figure}

The general mean-field Hamiltonian for a superconducting system
can be written in terms of the electron field operators 
$\hat{\psi}_{\sigma}(\vec{r})$

\begin{equation}
\hat{H} =  \int d\vec{r} \; \left\{ \sum_ {\sigma}
\hat{\psi}^{\dagger}_{\sigma} (\vec{r}) {\cal H}_e(\vec{r})
\hat{\psi}_{\sigma}(\vec{r}) + \Delta^*(\vec{r})
\hat{\psi}^{\dagger}_{\uparrow}(\vec{r})
\hat{\psi}^{\dagger}_{\downarrow}(\vec{r}) +
\Delta(\vec{r}) \hat{\psi}_{\downarrow}(\vec{r}) 
\hat{\psi}_{\uparrow}(\vec{r}) \right\}
\end{equation}

\noindent
where ${\cal H}_e$ is the one-electron Hamiltonian  and
$\Delta(\vec{r})$ is the superconducting order parameter. The problem of
calculating transport properties in such a continuous representation for
a non-homogenuous system is extremely involved requiring the knowledge of
the adequate boundary conditions at the interfaces.
Some attempts in this direction have
been recently presented by Zaitsev and Averin \cite{Zaitsev} within the
quasiclasssical Green functions approach.
A different approach which circumvent these difficulties while 
keeping a fully microscopic description of the problem can be obtained
by expanding the field operators in a discrete basis and writing the
Hamiltonian (1) in the form \cite{joseph}

\begin{equation}
\hat{H} =
\sum_{i ,\sigma} (\epsilon_{i} - \mu_i) c^{\dagger}_{i \sigma} c_{i
\sigma}
+ \sum_{i \neq j ,\sigma} t_{ij} c^{\dagger}_{i \sigma} c_{j \sigma} +
\sum_{i} (\Delta^{*}_{i} c^{\dagger}_{i \downarrow} c^{\dagger}_{i
\uparrow} + \Delta_{i} c_{i \uparrow} c_{i \downarrow}) ,
\end{equation}

\noindent
where $i,j$ run over the different sites of the system, $t_{ij}$ are the
hopping parameters connecting the different sites; $\mu_i$ and
$\Delta_i$ being the chemical potential and order parameter in a site
representation.
The simplification introduced by this approach allows to deal with rather
involved situations including spatial inhomogeneities (self-consistency)
and non-stationary effects typically appearing in superconductors.
For the voltage range $eV \sim \Delta$ the energy dependence of the
transmission coefficients can be neglected and the transport properties
can be expressed as a superposition of independent channels
\cite{Beenakker1}.
One can simplify even further the model to represent an SQPC with a
single conduction channel, which can be described by the 
following Hamiltonian 

\begin{equation}
\hat{H} = \hat{H}_{L} + \hat{H}_{R} +
 \sum_{\sigma} (t e^{i\phi(\tau)/2} c^{\dagger}_{L \sigma} c_{R \sigma} +
t^{*} e^{-i \phi(\tau)/2} c^{\dagger}_{R \sigma} c_{L \sigma})
- \mu_{L} \hat{N}_{L} - \mu_{R} \hat{N}_{R} ,
\end{equation}

\noindent
where $H_{L,R}$ are the BCS Hamiltonians for the left and right uncoupled
electrodes characterized by a constant order parameters $\Delta_{L,R}$
(for a symmetric contact $\Delta_L=\Delta_R=\Delta$). $\phi(\tau)$ 
is the time-dependent superconducting phase
difference entering as a phase factor in the hopping terms describing
electron transfer between the electrodes.
In our model the transmission, $\alpha$, can be varied
between 0 and 1 as a function of the coupling parameter $t$ (see
\cite{tocho} for details).
Within this model, the total current through the contact can be written
as

\begin{equation}
I(\tau)= \frac{i e}{\hbar}  \sum_{\sigma} \left( t e^{i
\phi(\tau)/2} <c^{\dagger}_{L\sigma}(\tau) c_{R \sigma}(\tau)> - \;
t^{*} e^{-i \phi(\tau)/2} <c^{\dagger}_{R \sigma}(\tau)
c_{L \sigma}(\tau)> \right) .
\end{equation}

The averaged quantities appearing in the current can be expressed in terms
of non-equilibrium Green functions \cite{Keldysh}. For
the description of the superconducting state it is useful to introduce spinor
field operators (Nambu representation) \cite{Nambu}, which in a site
representation are defined as

\begin{equation}
\hat{\psi_{i}} = \left(
\begin{array}{c}
c_{i \uparrow} \\ c^{\dagger}_{i \downarrow}
\end{array} \right) \hbox{  ,     } \hat{\psi}^{\dagger}_{i}=
\left(
\begin{array}{cc}
 c^{\dagger}_{i \uparrow} & c_{i \downarrow}
\end{array} \right) .
\end{equation}
\noindent
Then, the different correlation functions appearing in the Keldysh formalism
adopt the standard causal form

\begin{equation}
\hat{G}_{ij}^{\alpha,\beta}(\tau_{\alpha},\tau'_{\beta})=-i
< \hat{T}[\hat{\psi}_{i}(\tau_{\alpha})
\hat{\psi}_{i}^{\dagger}(\tau'_{\beta})]>
\end{equation}
\noindent
where $\hat{T}$ is the chronological ordering operator along the closed
time loop
contour \cite{Keldysh}. The labels $\alpha$ and $\beta$ refer to the upper
($\alpha \equiv +$)
and lower ($\alpha \equiv -$) branches on this contour.
The functions $\hat{G}^{+-}_{ij}$, which can be associated
within this formalism with
the electronic non-equilibrium distribution functions \cite{Kadanoff},
are given by the (2x2) matrix

\begin{equation}
\hat{G}^{+-}_{i,j}(\tau,\tau^{\prime})= i \left(
\begin{array}{cc}
<c^{\dagger}_{j \uparrow}(\tau^{\prime}) c_{i \uparrow}(\tau)>   &
<c_{j \downarrow}(\tau^{\prime}) c_{i \uparrow}(\tau)>  \\
<c^{\dagger}_{j \uparrow}(\tau^{\prime}) c^{\dagger}_{i \downarrow}(\tau)>  &
<c_{j \downarrow}(\tau^{\prime}) c^{\dagger}_{i \downarrow}(\tau)>
\end{array}  \right) .
\end{equation}

In terms of the $\hat{G}^{+-}$ the current is given by

\begin{equation}
I(\tau)  =  \frac{e}{\hbar} \; Tr \left[ \hat{\sigma}_z
\left( \hat{t} \hat{G}^{+-}_{RL}(\tau,\tau) -
\hat{t}^{\dagger} \hat{G}^{+-}_{LR}(\tau,\tau) \right) \right] ,
\end{equation}

\noindent
where $\hat{t}$ is the hopping element in the Nambu representation

\begin{equation}
\hat{t} = \left(
\begin{array}{cc}
 t   &   0      \\
  0                    &   -t^{*}
\end{array} \right)  .
\end{equation}

The Green functions 
$\hat{G}^{+-}_{ij}$ are calculated using an infinite order perturbation
theory with the coupling term in Eq. (3) considered as a perturbation.
Within this approach these Green functions obey a set of integral
Dyson equations \cite{tocho}. As discussed in the next sections, the
solution is strongly dependent on the biasing condition which determines
the time dependence in the superconducting phase difference.

\section{Current in a voltage biased contact}

The simplest biasing condition is that of a constant applied voltage.
This situation is rather easy to achieve experimentally except for very
small voltages (see section IV). In spite of its apparent simplicity
the theoretical analysis is quite complex because of the time-dependent
phase-difference which gives rise to a time dependent current containing
all harmonics of the Josephson frequency $\omega_0 = 2eV/\hbar$, i.e
$I(\tau) = \sum_n I_n(V) \exp{in\omega_0}$. The current can be also
decomposed into dissipative and nondissipative parts according to the
different symmetry with respect to $V$ of even and odd terms in the
previous expansion \cite{tocho}.

In this case the integral Dyson equations can be transformed into a set
of algebraic equations by a double Fourier transformation, defined by

\begin{equation}
\hat{G}_{n,m}(\omega) = \int d\tau \int d\tau^{\prime} 
e^{-i \omega_0 (n \tau - m \tau^{\prime})/2} e^{i \omega (\tau -
\tau^{\prime})} \hat{G}(\tau,\tau^{\prime})
\end{equation}

An efficient algorithm for the numerical evaluation of the Green 
function Fourier components is discussed in Ref. \cite{tocho}.

We shall concentrate in this section in the dc component of the current
$I_0$ which is the quantity more readily accessible experimentally.
Fig. 2 shows the dc $IV$ characteristics calculated from the fully
quantum mechanical theory and from the semiclassical OBTK theory
\cite{OBTK}. As can be observed the results become increasingly different 
for decreasing transmission. The fully quantum-mechanical calculation
exhibits a pronounced subgap
structure with steps at $eV = 2\Delta/n$ which is hardly noticeable in
the semiclassical theory. Both theories give the same result
nevertheless for perfect transmission where interference effects, not
included in the semiclassical theory, disappear due to the absence of
backscattering.

The experimental $IV$ characteristics for atomic contacts of different
metals are in remarkable agreement with our theoretical results. This is
illustrated in Fig. 3 for the case of a one-atom contact made of Pb
(these results are taken from Ref. \cite{Scheer2}).
This agreement has allowed to extract information on
the conduction channels transmissions $T_n$ of metallic atomic contacts
\cite{Scheer1,Scheer2,Cuevas2}.

\begin{figure}
\centering
\leavevmode
\epsfysize=5cm
\epsfbox{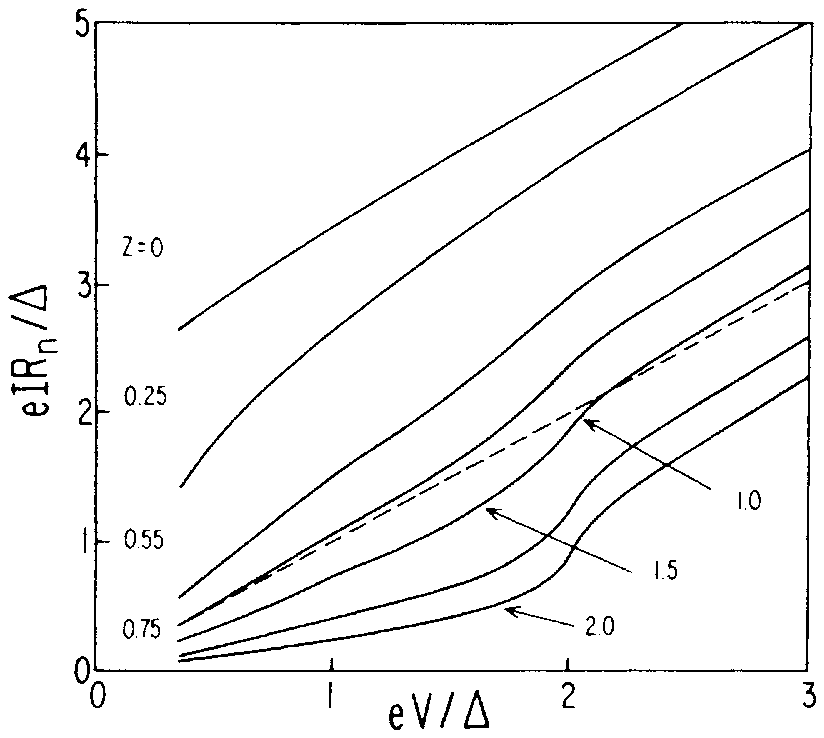}
\epsfysize=5cm
\epsfbox{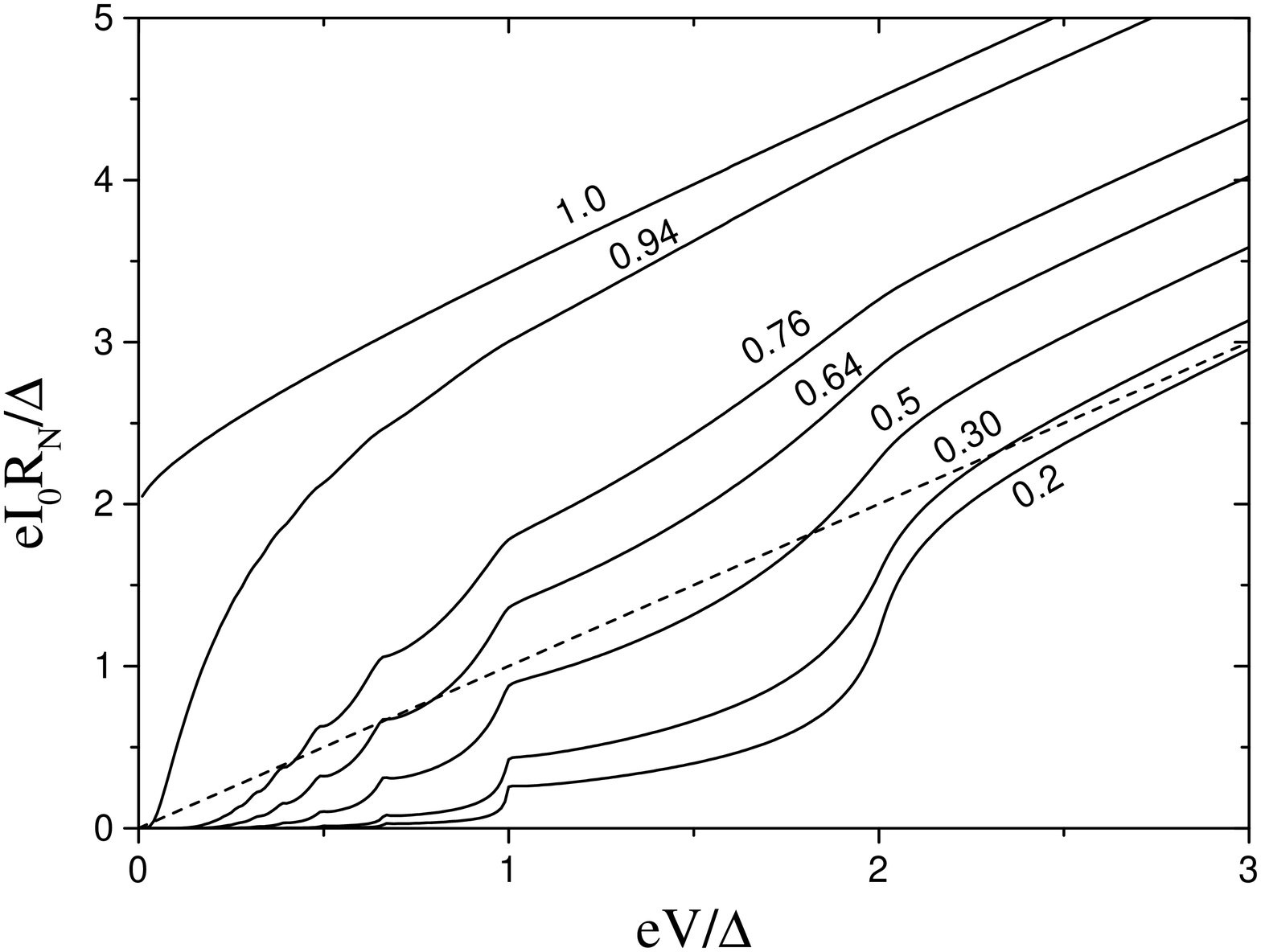}
\caption{dc current-voltage characteristics of a SQPC
for different values of the
normal transmission. Left panel corresponds to the semiclassical OTBK
theory and right panel to the fully quantum mechanical calculation.}
\end{figure}

\begin{figure}[!htb]
\begin{center}
\vspace{-0.5cm}
\epsfysize=5 cm
\epsfbox{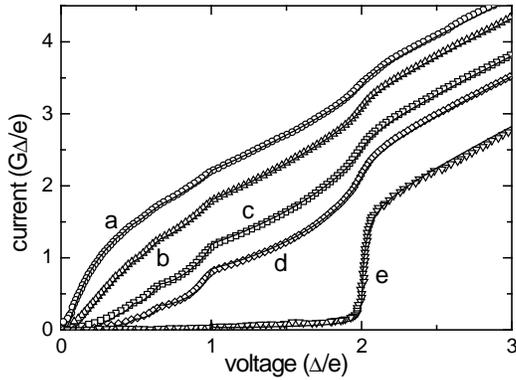}
\caption{Measured current-voltage characteristics (symbols) for
different realizations of a Pb one-atom contact at 1.5 K fabricated with
the STM technique [6]. The full lines are numerical fits
obtained by superposing four one-channel $IV$ curves with different 
transmissions.} 
\end{center}
\end{figure}

The temperature dependence of the $IV$ characteristics is shown in Fig.
4 for different values of the transmission. A remarkable feature of this
dependence is that the SGS persists up to temperatures close to the
critical temperature. When normalized to the
temperature dependent superconducting gap the dc current exhibits a
certain increase at low transmission, the opposite behavior being found
close to prefect transmission. The crossover between these two
tendencies is found for $\alpha \sim 0.8$. 

\begin{figure}[!htb]
\begin{center}
\hspace*{1cm}
\epsfxsize=9cm
\epsfbox{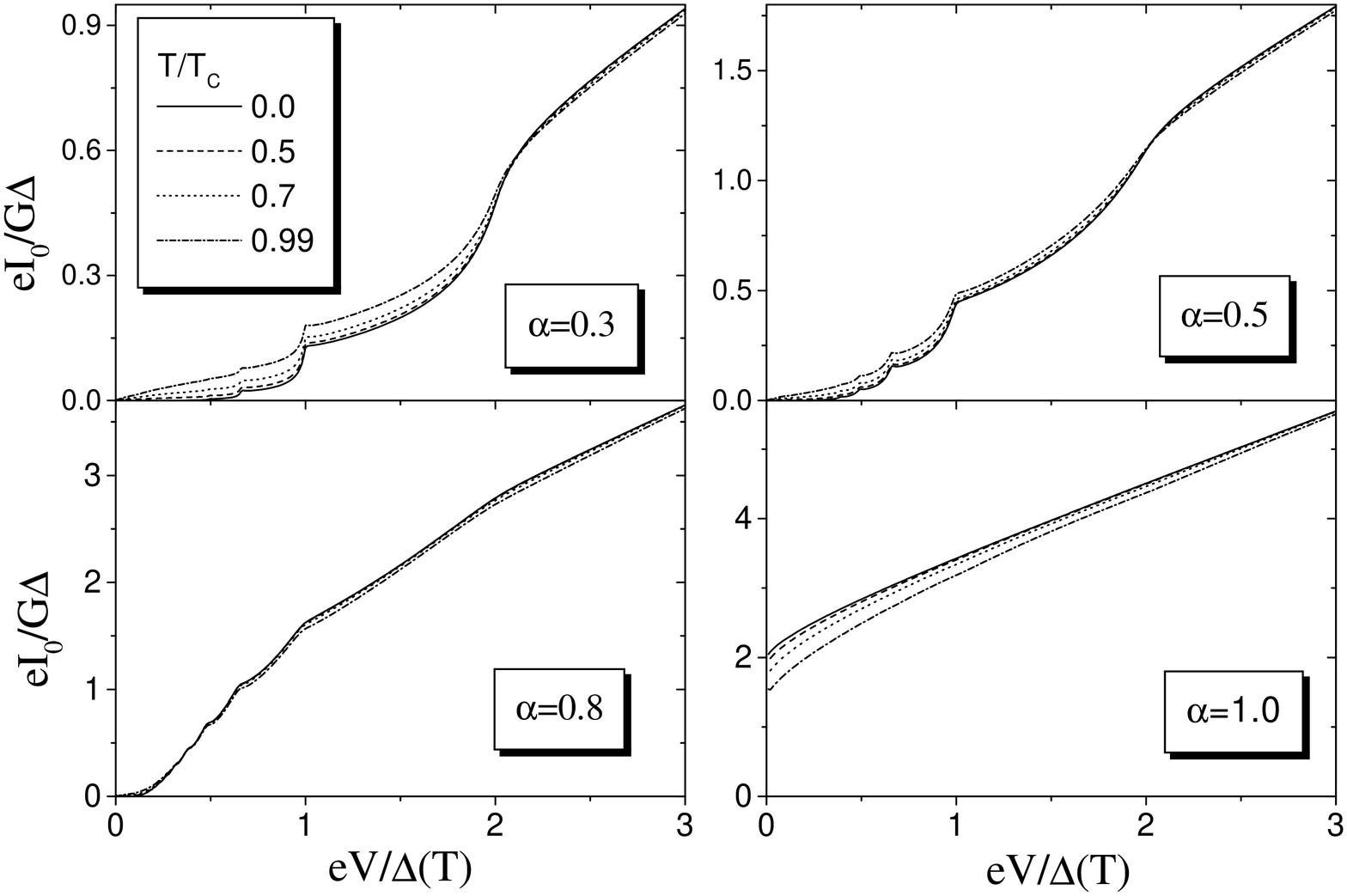}
\caption{dc current-voltage characteristic for different temperatures
and four values of the transmission.}
\end{center}
\end{figure}

The limit of very small bias is particularly interesting due to the
contribution of MAR processes of increasing order  $n \sim
2\Delta/eV$. This divergency in the $V \rightarrow 0$ limit is
eventually controlled by the presence of an inelastic relaxation rate
$\eta$ (usually a small fraction of the gap parameter) which introduces a
cut-off in the theory when $eV < \eta$. The effect of the inelastic
relaxation rate is to damp MAR processes of order $n > 2\Delta/\eta$.
As a consequence the system experiments a transition into a different
regime where the total current becomes linear in $V$. In this regime the
system response is determined by the adiabatic dynamics of the Andreev
states at $\epsilon(\phi) = \Delta \sqrt{1 - T \sin^2(\phi/2)} $ which
move following the actual value of the superconducting phase. The total
current can then be written as 
$I(\phi,V) = I_S(\phi) + G(\phi)V$ \cite{gphi}, where the supercurrent
$I_S(\phi)$ and the phase-dependent linear conductance $G(\phi)$
are given by

\begin{eqnarray}
I_S(\phi) & = &  \frac{e \Delta}{2 \hbar}
\frac{\alpha \sin \phi}{\sqrt{1-\alpha \sin^2(\phi/2)} }
\tanh (\frac{\beta \epsilon(\phi)}{2} )
\nonumber \\
G(\phi) & = & \frac{2e^2}{h} \frac{\pi}{16 \eta} \left[ \frac{\Delta
\alpha
 \sin \phi}{\sqrt{1-\alpha \sin^2 (\phi/2)} } \mbox{sech}
(\frac{\beta \epsilon(\phi)} {2}) \right]^2 \beta V .
\end{eqnarray}

The expression of the supercurrent \cite{joseph,RDA,Beenakker2} in Eq. (11)
interpolates between the Josephson $I_S \sim \sin{\phi}$ behavior and the
Kulik-Omelyanchuk $I_S \sim \sin{\phi/2}$ ballistic limit \cite{KO}.
This behavior
at high transmission has been recently confirmed experimentally using
break junction techniques in an SQUID configuration \cite{Koops}.
It should be noticed that the expression of $G(\phi)$
gives a definite answer to an old-standing problem
concerning the form  of this term known as the ``$cos{\phi}$ problem"
\cite{Barone}. The precise form of this term remains to be explored
experimentally (a similar set up to that used in Ref. \cite{Koops} could
be used for this purpose).

Finally, it should be stressed that from a mathematically point of view
the two limits $\eta \rightarrow 0$ and $V \rightarrow 0$ are not
interchangeable \cite{tocho}. In practice the limit $eV
\rightarrow 0$ with $eV > \eta$ can never be reached as there is always
a finite although small inelastic relaxation rate present.

\section{Current biased contact}

At very low voltages (and specially for high transmission) the contact
impedance may become actually smaller than the voltage source impedance.
If these conditions apply, the assumption of having an ideal source
providing a constant voltage bias which fixes the phase dynamics is no
longer valid. In this case one should take into account the
electromagnetic environment of the contact in order to determine the
phase dynamics and the system response to the external bias.

For conventional tunnel junctions this limit is usually analyzed by
means of the RSJ and RSCJ models \cite{Tinkham,Barone} which represent
the actual environment by a simple shunted circuit with a resistance $R$
and a capacitance $C$ connected in parallel to the junction. Within these 
simple
models the phase dynamics is equivalent to that of a particle moving in
a ``tilted washboard'' potential $U(\phi) = -I_b \phi + I_c \cos{\phi}$,
where $I_b$ is the biasing current and $I_c$ is the Josephson critical
current. At finite temperatures one should also consider thermal
fluctuations acting as an stochastic force on the fictitious particle.

To analyze the response of an SQPC under current bias one should
generalize these models for contacts of arbitrary transmission
\cite{shuntedAverin}. The
description of the superconducting phase as classical variable will be
valid as long as the Josephson coupling energy $E_J \sim \hbar I_c/2e$ is
much larger than the charging energy $E_C \sim e^2/2C$. For an SQPC
connected to a current source, 
the equations for the generalized RSCJ model would be given by

\begin{eqnarray}
I_b & = & \frac{\hbar}{2e} C \ddot{\phi} + \frac{\hbar}{2e} G(\phi)
\dot{\phi} + I_S(\phi) + i_n(\phi)
 \nonumber \\
V   & = & \frac{\hbar}{2e} \dot{\phi}
\end{eqnarray}

\noindent
where $I_S(\phi)$ and $G(\phi)$ are given by Eq. (11) of the previous
section and $i_n(\phi)$ is a fluctuating current whose power spectrum
$S$ is related to $G(\phi)$ by the fluctuation-dissipation theorem $S =
4k_BTG$ (see section VI). It should be noticed that the above equations
are strictly valid in the limit of small voltages induced on the
contact, i.e. $eV < \eta$, which is the condition for the validity of
Eq. (11) in section III. The actual value of $\eta$ is unknown but can
be estimated to be of the order of $\Delta/100$ or less \cite{comment}.
In the mechanical analogy, the effective potential for the generalized
RSCJ model can be written as

\begin{equation}
U(\phi) = - \left\{ I_b \phi + \frac{4e}{\beta \hbar} \log \left[
\mbox{cosh} \left( \frac{\beta \epsilon(\phi)}{2} \right) \right]
\right\}
\end{equation}

\noindent
and there appears a ``position'' dependent friction which comes from the
dissipative term in $G(\phi)$. The inclusion of this phase-dependent
term should have important consequences in the dynamics of the system.
Notice that the particular form of $G(\phi)$ (Eq. 11) introduces a
very asymmetrical friction with a minimum at the local minima of
$U(\phi)$ and maximum at the local maxima.

Integrating the Eqs. (12) of the generalized RSCJ model for arbitrary
conditions is a formidable task. An approximate solution for the
overdamped case, i.e. $G(\phi)/C > (2eI_c/\hbar C)^{1/2}$, can be
obtained following the procedure introduced by Ambegaokar and Halperin
\cite{AH} for overdamped tunnel junctions. The generalization of the
Ambegaokar and Halperin theory is straightforward once we have
identified the generalized potential (Eq. (13)) and the shunted
resistance with $1/G(\phi)$ (details will be given elsewhere).
The measurement of the slope of the $IV$ curve at zero voltage, which is
directly related to $G(\phi)$, would provide information on the 
value of $\eta$ in real systems.

\section{Contact under microwave radiation}

As discussed in section III,
the Andreev states play a central role in determining the adiabatic
dynamics of an SQPC at low bias voltage. Considering that typical
subgap energies are in the microwave range, it seems natural
to propose using microwave radiation for a
direct detection of Andreev states. This possibility has been suggested
in a previous work by us \cite{photo} and in Ref. \cite{Shumeiko2}.

The effect of a microwave external field can be easily introduced in the
single channel contact model. One can assume that the field intensity is
maximum in the constriction region and neglect the effect of field
penetrating inside the electrodes.
Within this assumptions the field can be introduced as a
phase factor modulating the hopping term $t$ in Eq. (3), i.e.

\begin{equation}
t(\tau) = t e^{i \alpha_0 \cos{\omega_r \tau}}, 
\end{equation}

\noindent
where $\omega_r$ is the microwave frequency, $\alpha_0 = eV_{opt}/(\hbar
\omega_r)$, $V_{opt}$ being the optical voltage induced by the field
across the constriction. The parameter $\alpha_0$ measures the strength
of the coupling with the external field.
The time-dependent hopping term can be expanded as

\begin{equation}
t(\tau) = t \sum_n i^n J_n(\alpha_0) e^{i n \omega_r \tau} ,
\end{equation}

\noindent
where $J_n$ is the $n$-order Bessel function.
For small coupling one can keep the lowest order terms in Eq. (15)
and obtain some analytical results \cite{photo}. 
In the general case, the model Hamiltonian can be viewed, according to Eq.
(15), as a superposition of processes where an arbitrary number of 
quanta of energy $\hbar \omega_r$ are absorbed or emitted.  
As the temporal dependence of each term in Eq. (15) is formally
equivalent to that in the constant voltage case,   
the generalization of the algorithm discussed
in section III to the present case is straightforward.

\begin{figure}[!htb]
\begin{center}
\epsfxsize=5cm
\epsfbox{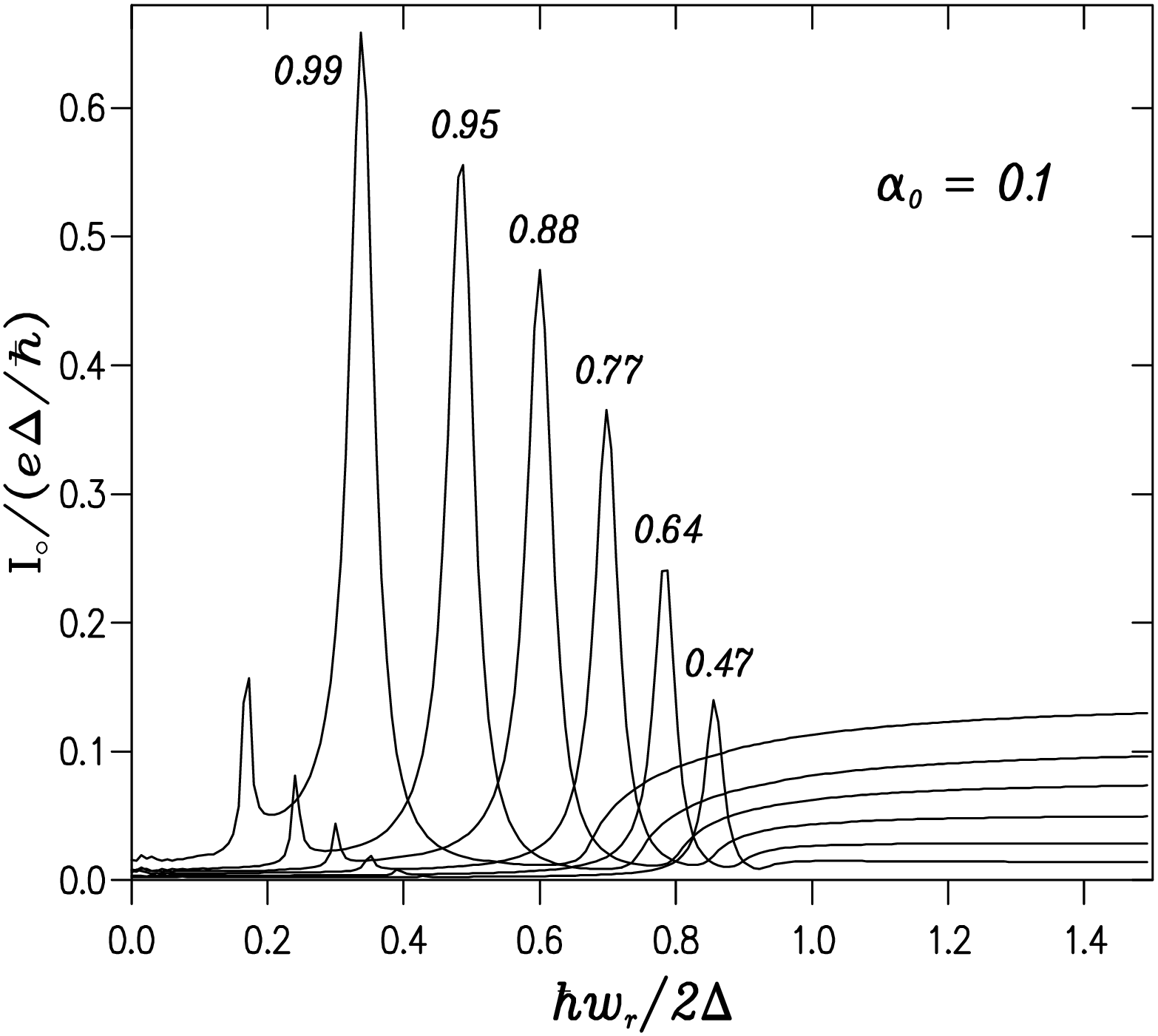}
\caption{Induced dc current in a SQPC under microwave-radiation for
different values of the transmission. The parameter $\alpha_0$ controls
the coupling to the external field (see text).}
\end{center}
\end{figure}

Fig. 5 shows the induced dc current as a function of the microwave
frequency for the case of low coupling constant ($\alpha_0 = 0.1$). 
All these results correspond to the situation in which the contact is
carrying the maximum supercurrent. In this weak coupling limit the
induced current is mainly due to the excitation from the lower to
the upper Andreev
state, which carries a negative current (i.e. opposite to the 
supercurrent). As a consequence the induced current exhibits a maximum
for the resonant condition $\omega_r = 2 \epsilon(\phi)$. At resonance,
the induced current can be of the same order as the critical supercurrent.
One can also notice a second stellite peak around $\epsilon(\phi)$ 
associated with {\it two photon} processes and a continuous band above
$\Delta + \epsilon(\phi)$. When the coupling constant $\alpha_0$
increases the contribution of higher order processes becomes
progresively more important giving rise to a complex structure where the
resonant condition for the excitation of the upper Andreev state can
no longer be resolved \cite{photo}.

\section{Thermal and shot noise}

The analysis of current fluctuations has a central role in the theory of
transport in mesoscopic systems \cite{deJong}. Fluctuations can provide
useful information on the microscopic dynamics (correlations) not
contained in the average current. It is thus desirable to develop a
fully quantum mechanical theory of current fluctuations in an SQPC
on an equal footing as previously discussed for the current.
In this respect, some limiting cases have already been analyzed in the
recent literature: the excess noise for $eV \gg \Delta$ in a ballistic
contact has been discussed in Ref. \cite{Shuminoise}, thermal noise
for arbitrary transmission was analyzed in Ref. \cite{thermal} while 
the case of perfect transmission and finite voltages has been addressed
in Ref. \cite{Averinnoise}.

The noise power spectrum is defined by

\begin{equation}
S(\omega,\tau) =  \hbar \int d \tau^{\prime} \; e^{iw \tau^{\prime}}
<\delta \hat{I}(\tau + \tau^{\prime}) \delta \hat{I}(\tau)
+ \delta \hat{I}(\tau) \delta\hat{I}(\tau + \tau^{\prime})>
\end{equation}

\noindent
where $\delta \hat{I}(\tau) =\hat{I}(\tau) - <\hat{I}(\tau)>$. 
For the evaluation of the above
correlation functions a BCS mean field decoupling procedure can be
performed. $S(\omega,\tau)$ can then be written in terms of
nonequilibrium Green functions introduced in section III. In the voltage
biased case, $S(\omega,\tau)$ can be expanded in harmonics of the
Josephson frequency, i.e. $S(\omega,\tau) = \sum S_n(\omega)
\exp{in\omega_0 \tau}$. As in the case of the average current the noise
Fourier components $S_n(\omega)$ can be evaluated in terms of the Green
functions matrix elements $G_{n,m}$ defined in section III.

Let us start by analyzing the $V = 0$ case where noise is due to thermal
fluctuations. While in a normal QPC thermal noise has the usual well
understood behavior, increasing
linearly with temperature and with a flat frequency spectrum, in the
superconducting case it exhibits very unusual behavior as a function of
temperature, frequency and phase. Fig. 6 illustrates the frequency
dependence of the thermal noise for different transmissions. As can be
observed the noise exhibits two sharp resonances at $\omega=0$ and
$\omega = 2 \epsilon(\phi)$ corresponding to the excitation of the
Andreev bound states. For $\omega > \Delta + |\epsilon(\phi)|$, $S$
exhibits a broad band arising from the continuous part of the
single particle spectral density. 

\begin{figure}[!htb]
\begin{center}
\epsfxsize=5cm
\epsfbox{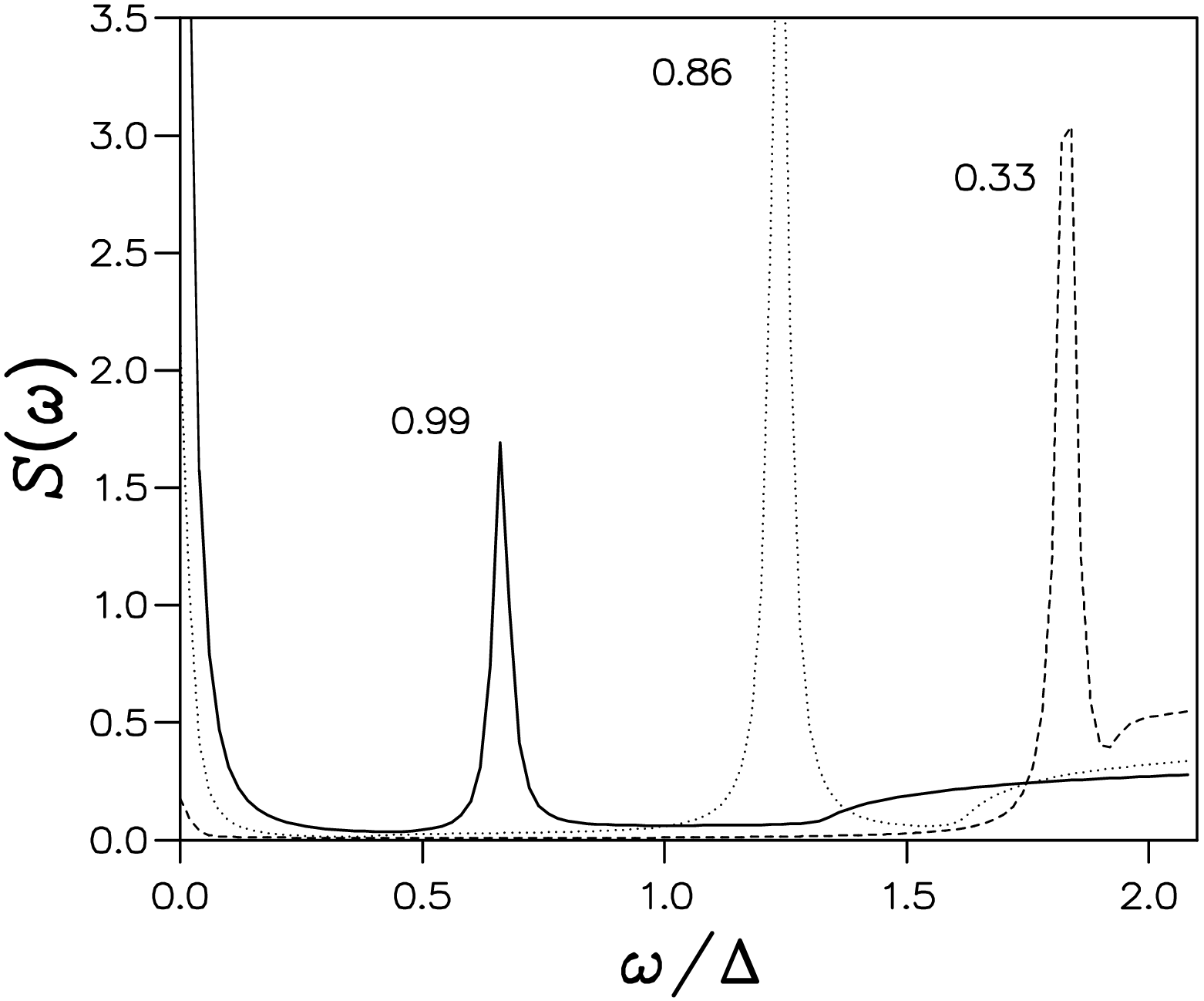}
\caption{Power spectrum $S(\omega)$ of the zero-voltage
current-fluctuations at $k_BT = 0.2 \Delta$ in a SQPC for three 
transmission values. The
contact is biased at the maximum supercurrent.}
\end{center}
\end{figure}

The weight of the peaks at $0$ and $2 \epsilon(\phi)$ can be evaluated
analytically as discussed in Ref. \cite{thermal}. We find that the zero 
frequency noise is related to the phase-dependent linear conductance by
$S(0) = 4 k_B T G(\phi)$ as expected from the fluctuation dissipation
theorem. The exponential temperature dependence in $G(\phi)$ gives rise
to an exponential increase of thermal noise when $k_BT \sim
\epsilon(\phi)$. It should be noticed that the ratio $S(0)/2eI_S(\phi)$
can actually be divergent for any temperature provided that $\alpha
\rightarrow 1$ and $\phi \rightarrow \pi$. 
On the other hand, the weight of the peak at $2 \epsilon(\phi)$  is zero
for perfect transmission, increasing as $\alpha^2(1 - \alpha)$,  
becoming the dominant feature for finite $1 - \alpha$ and sufficiently
low temperatures. In fact, the ratio between $S(2 \epsilon(\phi))$ and
$S(0)$ is given by

\begin{equation}
S(2 \epsilon(\phi))/S(0) = \frac{1}{2}(1 - \alpha) \tan^2{\frac{\phi}{2}}
\mbox{cosh}\left[\frac{\epsilon(\phi)}{k_BT}\right] .
\end{equation}

Another quantity which is interesting to analyze and is directly
amenable to experimental measurement is the shot-noise. Mathematically
the shot-noise is given by the zero frequency dc component in the expansion 
of the noise power spectrum, i.e $S_0(0)$, at $eV \gg k_BT$. For
simplicity we will consider the zero temperature case.  Results for the
shot-noise as a function of voltage are shown in Fig. 7 for several
transmissions. The curves exhibit a pronounced subgap structure at the
voltage values $eV = 2\Delta/n$ as in the dc current. In the case of the
shot-noise the structure is more pronounced and is still 
observable for transmissions rather close to 1. In the perfect
ballistic limit shot-noise is greatly reduced 
due to correlations associated with the Pauli principle  
as in the case of a normal ballistic contact \cite{Khlus}.  

On the other hand, in the tunnel limit the shot-noise is expected to
reach the Poisson limit $S \sim 2 q I$, where $q$ is the transmitted
charge in an elementary process. This relation offers the possibility to
directly check whether multiple charges $q = ne$ are actually being
transmitted coherently in a n-th order MAR process \cite{Dieleman}.
Our theory allows to calculate the effective charges defined by the
shot-noise current ratio. In the tunnel limit one finds that $q$
exhibits a well defined step like behavior $q/e = 1 + Int[2\Delta/eV]$
confirming the above hypothesis \cite{Cuevas}. 

\begin{figure}[!htb]
\begin{center}
\hspace*{1cm}
\epsfysize=5cm
\epsfbox{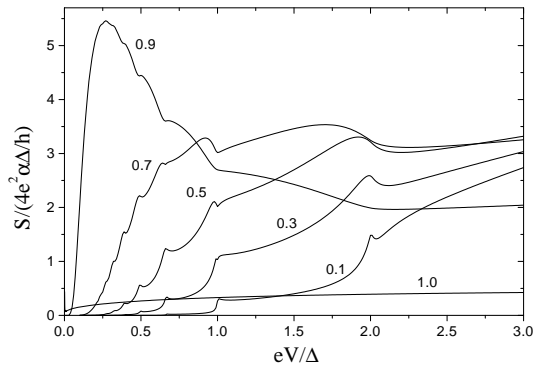}
\caption{Zero-frequency current-fluctuations at zero temperature
(shot-noise) as a function of bias voltage.}
\end{center}
\end{figure}

\section{SGS and resonant tunneling}

The model discussed so far describes an SQPC with an energy independent
transmission $\alpha$. In some situations, that can be achieved
experimentally, the normal transmission can have a non-negligible
variation on an energy scale of the order of $\Delta$. This can happen
when the constriction region is weakly coupled to the electrodes by
tunnel barriers as in the case of a small metallic particle or a quantum
dot coupled to superconducting leads \cite{Ralph}. 
We represent this situation by  the following model Hamiltonian

\begin{equation}
\hat{H}= \hat{H}_L + \hat{H}_R + \sum_{\nu,\sigma} t_{\nu} 
(\hat{c}^{\dagger}_{\nu \sigma} \hat{c}_{0 \sigma} +
\hat{c}^{\dagger}_{0 \sigma} \hat{c}_{\nu \sigma}) +
\sum_{\sigma} \epsilon_0 \hat{n}_{0 \sigma} +
U \hat{n}_{0 \uparrow} \hat{n}_{0 \downarrow}
\end{equation}

\noindent
where $\hat{H}_L$ and $\hat{H}_R$ describe the left and right leads,
$\epsilon_0$ is a resonant level associated with the isolated
constriction region, 
$t_{\nu}$ with $\nu = L, R$ are 
hopping parameters which connect the level to the left and
right leads. The $U$ term takes into account the 
Coulomb repulsion in the constriction region. 
The parameter $U$ is basically the  
charging energy, $E_C$, and is related to the central region capacitance 
$C$, by $U \sim e^2/2C$.
For the subsequent discussion it is convenient to introduce the normal 
elastic
tunneling rates $\Gamma_{\nu} = \pi  |t_{\nu}|^2 \rho_{\nu}(\mu)$,
where $\rho_{\nu}(\mu)$ are the normal spectral densities of the
leads at the Fermi level.

When the charging energy is much larger than both $\Gamma$ and $\Delta$
Andreev reflections are completely suppressed and transport is only
due to single-quasiparticle tunneling. This situation has been achieved
in experiments on transport through nanometer metallic particles
by Ralph et al. \cite{Ralph}. Model calculations presented by us 
in Ref. \cite{level} based on Hamiltonian (18) yield good agreement with
the experimental results. 

We shall consider in more detail the case of small charging energy, in
which the interplay between resonant tunneling and MAR give rise to 
novel effects and a very rich subgap structure \cite{level,Johansson}. 
Fig. 8 shows the dc $IV$ characteristic for different positions of the
resonant level $\epsilon_0$ with respect to the Fermi level. The
tunneling rates are taken in this case as $\Gamma_L = \Gamma_R = 
\Delta$. As can be observed, when the level is far from the gap region 
(case a) the limit of energy independent transmission is recovered and
the subgap structure is similar to the one depicted in Fig. 2 (right
panel). As the resonant level approaches the gap region the subgap structure
becomes progressively distorted with respect to the energy independent
transmission case. While the structure corresponding to the opening of
odd-order MAR processes (i.e. at $eV \sim 2\Delta/n$ with odd $n$)
is enhanced, the structure at $eV \sim 2\Delta/n$ with even $n$ is
suppressed. 
When $\Gamma \rightarrow 0$ (not shown), one can also notice the 
appearance of resonant peaks in the $IV$ characteristic for 
$eV \sim 2 \epsilon_0$. 

\begin{figure}[!htb]
\begin{center}
\hspace*{1cm}
\epsfysize=5 cm
\epsfbox{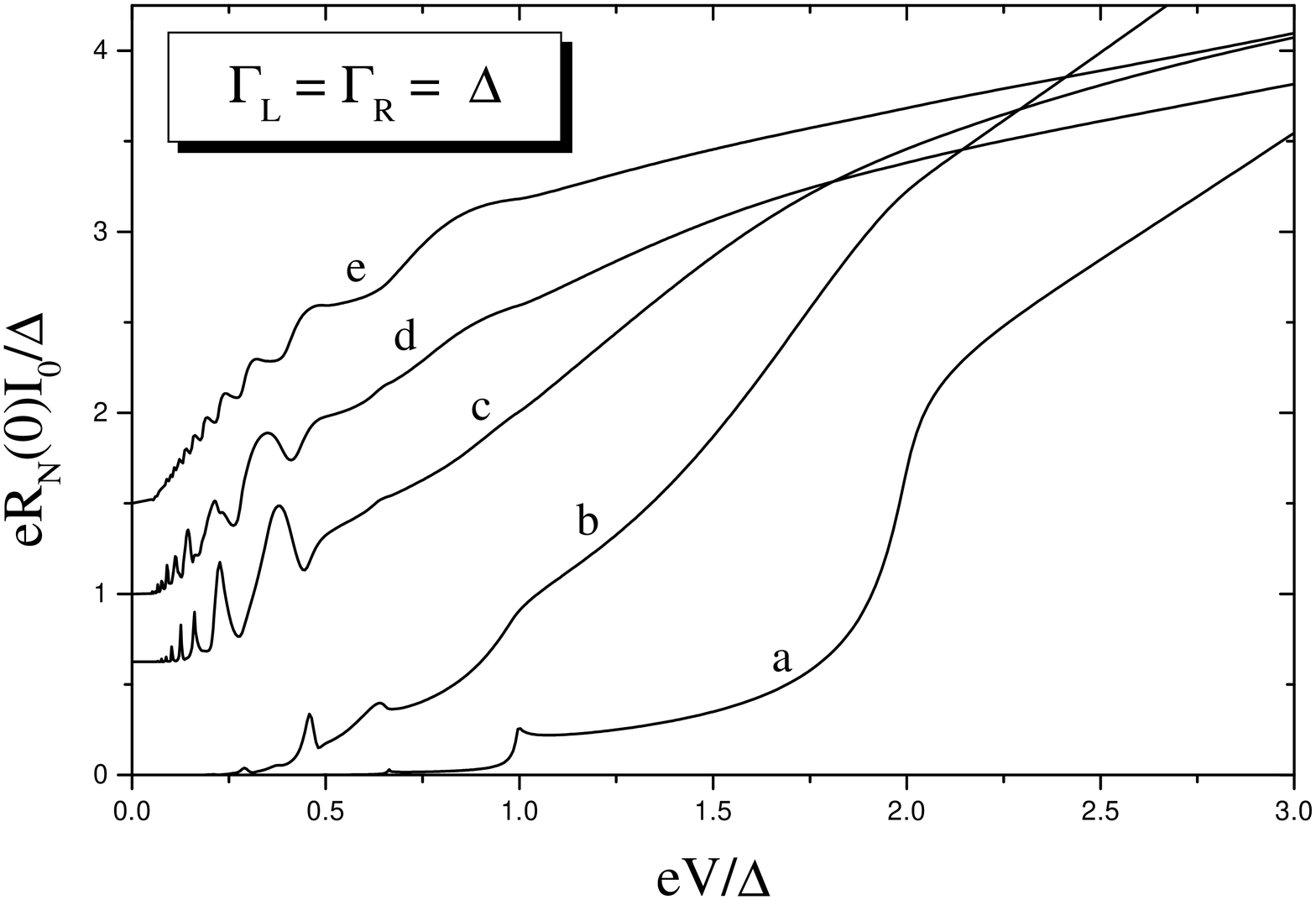}
\caption{dc current-voltage characteristic for a resonant level 
$\epsilon_0$ coupled
to superconducting leads. $\epsilon_0 = 5 \Delta$ (a), 
$2 \Delta$ (b), $\Delta$ (c), $\Delta/2$ (d) and 0 (e). $R_N(0)$ is the
normal resistance at the Fermi level. Curves (c), (d) and (e) have been
displaced for the sake of clarity.}
\end{center}
\end{figure}

\section{conclusions}

An overview of the results of a microscopic theory for the transport
properties of superconducting quantum point contacts has been presented.
These results include the response under different biasing conditions
for an energy independent transmission as well as the case of
resonant transmission.
A remarkable agreement has been found between the calculated
and the experimental dc $IV$ curves for atomic-size contacts 
\cite{Scheer1,Scheer2}. The agreement has allowed to extract information on
the number and transmissions of the conduction channels in
atomic-contacts of different metallic elements \cite{Scheer2,Cuevas2}. 
On the other hand, the agreement shows the importance of interference
effects included in a fully quantum-mechanical calculation and thus the
need to go beyond semiclassical theories for describing this kind of 
systems. 
Additional predictions of the microscopic theory remain to be
analyzed experimentally. For instance, we could point out the phase
dependence of the linear conductance, the direct observation of Andreev
levels in contacts under microwave radiation and the analysis of
shot-noise. This last analysis would provide direct evidence of 
coherent transmission of multiple charges in MAR processes.
Finally, the rich SGS in the presence of resonant transmission could
be explored in S-2DEG-S devices wich are currently being developed
\cite{Takayanagi}. 

\acknowledgements 
The authors would like to thank C. Urbina for fruitful discussions.
This work has been supported by the Spanish CICyT under contract
PB97-0044.


\begin{references}
\bibitem{comment} This small value is consistent with the agreement
between theory and experiments in the constant voltage case.
\end{references}
\end{document}